\documentclass{article}
\usepackage{amsmath, amssymb}
\usepackage[utf8]{inputenc}
\usepackage[T1]{fontenc}
\usepackage[english]{babel} 
\usepackage{float}
\usepackage{graphicx}
\usepackage{PRIMEarxiv}
\usepackage{booktabs}
\usepackage{url}
\usepackage{fancyhdr}
\usepackage{newunicodechar}
\newunicodechar{≤}{\leq}
\usepackage{pgfplots}
\pgfplotsset{compat=1.17}
\usepackage{algorithm}
\usepackage{algorithmic}
\usepackage{hyperref}
\usepackage{cleveref}

\title{Statistical Analysis of the Depth-Velocity Trade-off in Reflection Seismology}

\author{
    Rafael da Silva Garcia \\
    Federal University of Rio Grande do Norte (UFRN) \\
    Bachelor in Information Systems \\
    \texttt{garciarafael.2298@gmail.com} \\
    \And
  Francisco Márcio Barboza \\
Department of Computing and Technology \\
Federal University of Rio Grande do Norte \\
Caicó, Brazil \\
    \texttt{marcio.barboza@ufrn.br}   
}

\begin{document}
\maketitle

\begin{abstract}
Seismic interpretation is strongly influenced by the relationship between subsurface layer depth and velocity. Small variations in these parameters can produce almost identical responses, characterizing the depth-velocity trade-off phenomenon. This work proposes a statistical and computational approach to evaluate the extent of this effect through Monte Carlo simulations of thousands of synthetic models. The analysis involves generating random depth–velocity pairs, computing travel times, and calculating the root-mean-square error (RMSE) relative to a base model. Results show that multiple parameter combinations yield nearly indistinguishable travel times, confirming the presence of wide ambiguity regions. Finally, a two-layer geological velocity model is presented to illustrate the structural–velocity relationship.
\end{abstract}
\vspace{0.5cm}

\keywords{Reflection Seismology \and Depth-Velocity Trade-off \and Ambiguity \and Python}

\section{Introduction}

Reflection seismology is an essential technique for characterizing the Earth's subsurface. Its basic principle involves measuring the travel times of seismic waves reflected at geological interfaces, allowing the estimation of layer depths and velocities \cite{yilmaz2001seismic,aki2002quantitative}.

However, this estimation is not unique: different combinations of depth ($z$) and velocity ($v$) can generate similar travel times, a phenomenon known as the Depth-Velocity \textit{trade-off}. Such ambiguity can lead to erroneous interpretations, especially when auxiliary data (such as well logs or seismic inversion) are unavailable \cite{dix1955seismic,stork1992reflection}.

Global inversion and optimization studies show that the resolution of the seismic inverse problem is often non-unique, making it necessary to incorporate probabilistic constraints and statistical modeling \cite{tarantola1987inverse,mosegaard1995monte,sen2013global}.

This work aims to statistically quantify the ambiguity associated with this \textit{trade-off}. Through random simulations and error analysis (RMSE), we evaluated the frequency and distribution of models that produce responses nearly indistinguishable from a base scenario.

\section{Methodology}

The proposed methodology is based on numerical simulations implemented in the Python language, widely used in geosciences due to its flexibility and robustness for data analysis and scientific visualization \cite{oliphant2007python,lin2012python}.

The analysis of temporal similarity is based on the root mean square error (RMSE), a concept widely used in seismic inversion studies and geological model fitting \cite{tarantola1987inverse, mosegaard1995monte}.

\subsection{Base Model} 

The base model consists of a horizontal reflector at 400 m depth and a velocity of 2000 m/s. The travel time is calculated according to:
\begin{equation}
    T(x) = \sqrt{\left(\frac{2z_0}{v_0}\right)^2 + \left(\frac{x}{v_0}\right)^2}
\end{equation}
where $x$ represents the offset between source and receiver.
This type of formulation is traditional in velocity analysis and forms the basis of the hyperbolic approximation used in seismic processing \cite{yilmaz2001seismic}.

\subsection{Random Model Generation}

5000 pairs of $(z, v)$ were generated with values uniformly distributed in the following intervals:
\begin{itemize}
    \item Depth: 200–600 m
    \item Velocity: 1000–3000 m/s
\end{itemize}

The $z/v$ ratio was used as an indicator of ambiguity, as models with values close to that of the base model ($z_0/v_0 = 0.2$ s) tend to generate equivalent travel times. This type of statistical approach reflects the concept of non-uniqueness in probabilistic inversion \cite{tarantola1987inverse}.

\subsection{RMSE Calculation}

For each model, the travel time $T_i(x)$ was calculated and compared with the travel time of the base model $T_0(x)$. The root mean square error is given by:
\begin{equation}
    RMSE = \sqrt{\frac{1}{N}\sum_{j=1}^N (T_i(x_j) - T_0(x_j))^2}
\end{equation}

Low RMSE values indicate ambiguous models, i.e., different $(z, v)$ pairs capable of reproducing the same observable response \cite{tarantola1987inverse}.

\subsection{Monte Carlo Simulation Algorithm}
To formalize the analysis process, the workflow is summarized in Algorithm \ref{alg:monte_carlo}. This pseudocode describes the iterative logic for model generation and error quantification, which is the basis of the Monte Carlo simulation.

\begin{algorithm}[H]
\caption{Monte Carlo Simulation for Trade-off Analysis} 
\label{alg:monte_carlo}
\begin{algorithmic}[1]
\REQUIRE $Z_{\text{range}}$, $V_{\text{range}}$, $N_{\text{samples}}$, $X_{\text{vector}}$, $Z_0$, $V_0$.
\ENSURE $RMSE_{\text{vector}}$
\STATE Calculate $T_{\text{base}}$ using $Z_0$, $V_0$ over $X_{\text{vector}}$ (Eq. 1).
\FOR{$i = 1$ \TO $N_{\text{samples}}$}
    \STATE $V_i \gets$ Generate uniform number in $V_{\text{range}}$.
    \STATE $Z_i \gets$ Generate uniform number in $Z_{\text{range}}$.
    \STATE $T_i \gets$ Calculate travel time with $Z_i$ and $V_i$ (Eq. 1).
    \STATE $\text{DIF} \gets T_i - T_{\text{base}}$.
    \STATE $RMSE_i \gets$ Calculate $RMSE(\text{DIF})$ (Eq. 2).
    \STATE Add $RMSE_i$ to the vector $RMSE_{\text{vector}}$.
\ENDFOR
\RETURN $RMSE_{\text{vector}}$
\end{algorithmic}
\end{algorithm}

\subsection{Computational Tools and Optimization} 
The implementation of Algorithm \ref{alg:monte_carlo} utilized the NumPy library for performance optimization. Instead of using the explicit iterative \textit{loop} described in the pseudocode, the actual code was rewritten using the \textbf{vectorization} and \textit{broadcasting} technique of NumPy. This optimization (seen in the Python code through the manipulation of $z_{\text{col}}$, $v_{\text{col}}$, and $x_{\text{row}}$ dimensions) allows the calculation of the travel time matrix ($T_{\text{models}}$) to be performed in a parallel and efficient manner. This design consideration was fundamental to ensure the speed and robustness of the Monte Carlo simulation with 5000 samples.

\section{Results and Discussion}

The results obtained from the Monte Carlo simulation not only confirm the existence of the depth-velocity \textit{trade-off} but also quantify the extent of the ambiguity region in the parameter space $(v,z)$, using the RMSE error as a metric of indistinguishability.

\subsection{Geological Model and Conceptual Ambiguity}

\begin{figure}[H]
    \centering
    \includegraphics[width=0.6\linewidth]{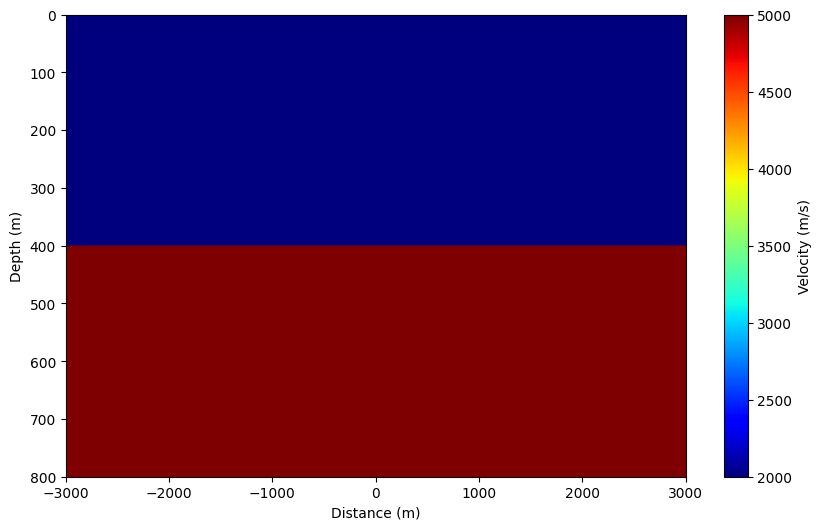}
    \caption{Synthetic two-layer geological model, with interface at 400 m.} 
    \label{fig:velocity_model}
\end{figure}

The synthetic Geological Model presented in \cref{fig:velocity_model} serves as a structural reference for the analysis. It is composed of a homogeneous upper layer of $2000\text{ m/s}$ (defining the $v_0$ of the Base Model) and a lower layer of $5000\text{ m/s}$, separated by the reflector at $400\text{ m}$. It is important to note that the \textit{trade-off} analysis in Section 3 is performed only on the parameters $(z,v)$ of the upper layer, as per Equation 1, since the velocity of the lower layer does not influence the reflection travel time in a single-layer formulation.

\begin{figure}[H]
    \centering
    \includegraphics[width=0.6\linewidth]{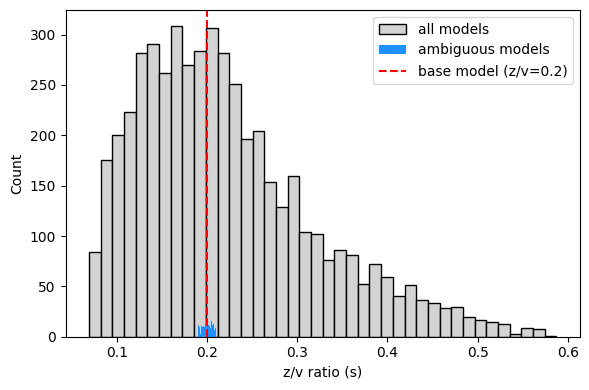}
    \caption{Distribution of the $z/v$ ratio with emphasis on ambiguous models close to 0.2 s.}
    \label{fig:zv_ratio}
\end{figure}

The $z/v$ ratio is a direct indicator of the travel time at zero offset ($T(0)$). Figure \cref{fig:zv_ratio} shows the distribution of this ratio. Models with $z/v$ between 0.19 and 0.21 (shaded region), which encompass the base value of $z_0/v_0 = 0.2$ s, were classified as potentially ambiguous, representing about 9\% of the samples.

\subsection{Parameter Space and the Nature of the Trade-off}

\begin{figure}[H]
    \centering
    \includegraphics[width=0.6\linewidth]{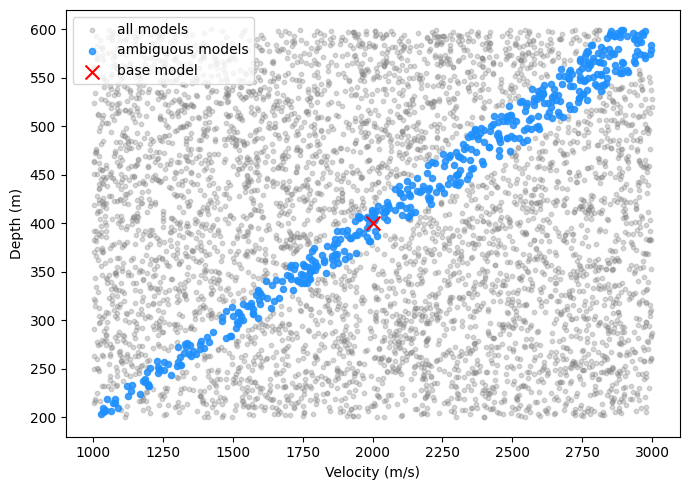}
    \caption{Distribution of the generated models in the $(v,z)$ space, highlighting the conceptual ambiguity region.} 
    \label{fig:vz_space}
\end{figure}

Figure \cref{fig:vz_space} presents the distribution of the tested models in the $(v,z)$ space. The blue band represents the ambiguous combinations according to the $z/v$ criterion, which form a nearly linear band around the ratio $z/v = 0.2$.

This distribution demonstrates that, to maintain an approximately constant reflection travel time, depth ($z$) and velocity ($v$) must be directly proportional, confirming the strong linear correlation inherent to the depth-velocity \textit{trade-off} \cite{dix1955seismic}.

\subsection{The Effect in the Time Domain: Hyperbola Overlap} 

\begin{figure}[H]
    \centering
    \includegraphics[width=0.6\linewidth]{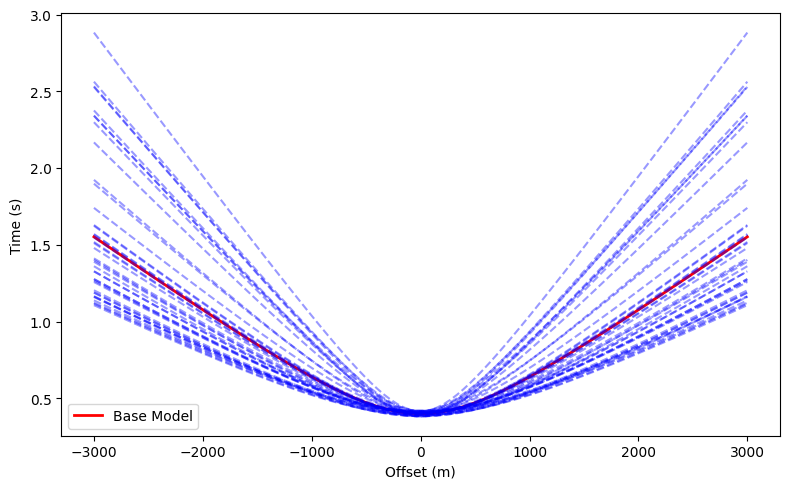}
    \caption{Hyperbola overlap of the base model and ambiguous models ($z/v$ criterion).}
    \label{fig:hyperbolas}
\end{figure}

Figure \cref{fig:hyperbolas} illustrates the impact of ambiguity in the observable domain. An overlap is observed between the hyperbolic curve of the base model (red) and dozens of ambiguous models (blue), demonstrating the difficulty in distinguishing the scenarios based only on the measured reflection times.

\subsection{Error Distribution and Indistinguishability Criterion} 

\begin{figure}[H]
    \centering
    \includegraphics[width=0.6\linewidth]{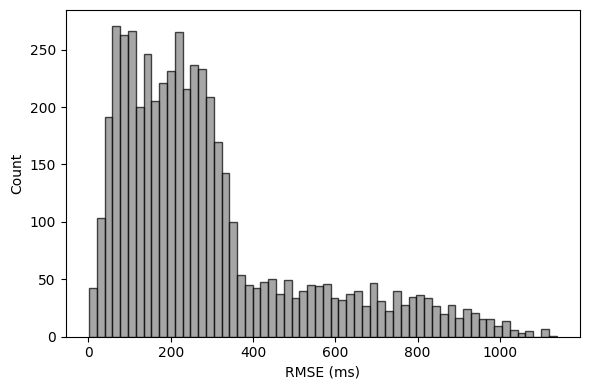}
    \caption{Distribution of RMSE relative to the base model. Small values indicate strong ambiguity.}
    \label{fig:rmse_hist}
\end{figure}

To quantify ambiguity more rigorously, we analyzed the distribution of the RMSE error. Figure \cref{fig:rmse_hist} indicates that the highest concentration of models has an RMSE below 400 ms, reinforcing that similar times are frequent over a wide range of the parameter space. This distribution confirms the existence of a vast ambiguity region.

We adopted a practical error limit of 10 ms (0.01 s), considered the typical noise level of a seismic survey, as the criterion for practical indistinguishability.

\begin{figure}[H]
    \centering
    \includegraphics[width=0.6\linewidth]{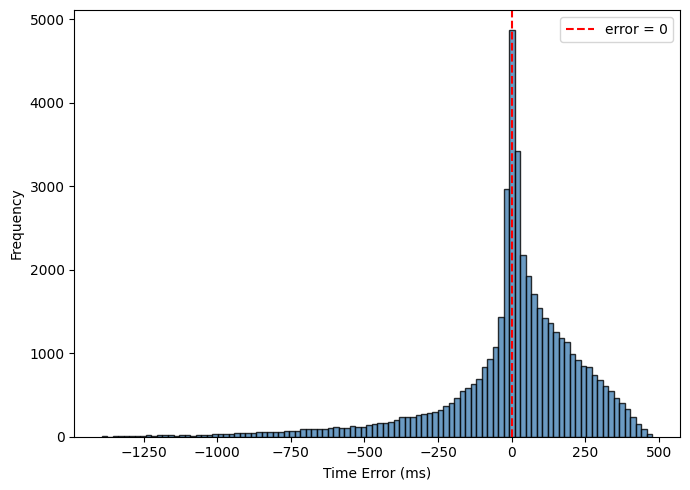}
    \caption{Distribution of point-to-point errors (in ms) for ambiguous models.}
    \label{fig:error_distribution}
\end{figure}

Figure \cref{fig:error_distribution} shows the distribution of point-to-point time errors. The sharp concentration of frequency around 0 ms (indicated by the dashed line) confirms the high temporal similarity at most offset points for the models classified as ambiguous.

\subsection{Detailed Analysis of Indistinguishable Models (RMSE $\le 10 \text{ ms}$)}

Within the $RMSE \leq 10 \text{ ms}$ criterion, 16 models (0.32\% of the total) were identified as practically indistinguishable from the base model. This result indicates that the region of strict ambiguity, within the limits of acquisition noise, is reduced but highly correlated with the characteristic band of the $z/v$ ratio.

\begin{figure}[H]
    \centering
    \includegraphics[width=0.6\linewidth]{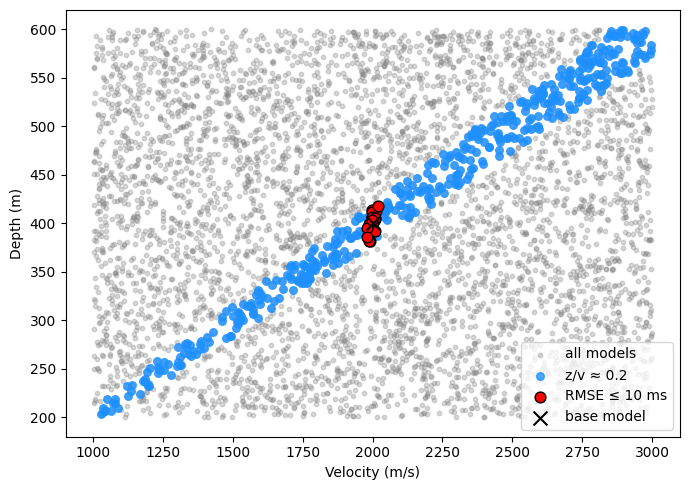}
    \caption{Distribution of models in the $(v,z)$ space, highlighting in red the cases with $RMSE \leq 10 \text{ ms}$.} 
    \label{fig:vz_rmse}
\end{figure}

Figure \cref{fig:vz_rmse} presents the complete distribution of the models, highlighting in red the 16 cases that satisfy the strict $RMSE \leq 10 \text{ ms}$ criterion. While the randomly generated models occupy a wide range on the plane, the ambiguous ones form a compact, almost perfectly linear region around the base model.

\begin{figure}[H]
    \centering
    \includegraphics[width=0.6\linewidth]{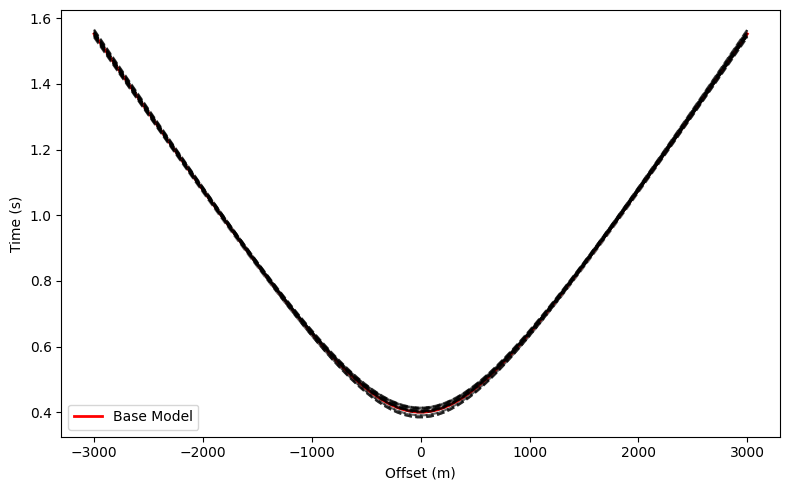}
    \caption{Overlap of the hyperbolas of the base model (black) and the 16 ambiguous models with $\mathbf{RMSE \leq 10 \text{ ms}}$. The curves practically coincide along the entire offset.} 
    \label{fig:hyperbolas_rmse}
\end{figure}

The strongest proof of indistinguishability is given in \cref{fig:hyperbolas_rmse}, which shows the hyperbolas of the 16 cases superimposed on the base model curve. The near-perfect coincidence along the entire offset reinforces that the space of viable solutions is continuous and follows a strong linear correlation between depth and velocity, in agreement with the principles of non-uniqueness discussed by Tarantola (1987) \cite{tarantola1987inverse}.

The obtained results corroborate previous observations by Stork (1992) regarding the difficulty of differentiating structures when variations in velocity and depth mutually compensate each other \cite{stork1992reflection}.

\section{Conclusion}

The present study quantitatively demonstrated the existence and extent of the depth-velocity \textit{trade-off} phenomenon. The statistical approach adopted showed that there is a large number of $(z,v)$ combinations capable of reproducing virtually identical reflection times.

The use of a practical error criterion ($RMSE \leq 10 \text{ ms}$) allowed the identification of strictly indistinguishable models, confirming a strong linear correlation between $z$ and $v$ for maintaining equivalent times.

However, it is important to emphasize that this ambiguity does not represent an exact mathematical non-uniqueness, but rather an interpretive uncertainty. Each $(z,v)$ pair generates a single travel time $T(x)$; however, different combinations of depth and velocity can produce temporal responses so similar that they become indistinguishable within the noise level and temporal resolution of the seismic data. This characteristic is inherent to the nature of seismic surveys and reflects the observational limitation, not a problem of multiple exact solutions.

Thus, the depth-velocity \textit{trade-off} is interpreted as a practical ambiguity, where small, compensatory variations in depth and velocity result in equivalent times within the precision of the experiment. This behavior confirms the principle of "non-uniqueness within the limits of data resolution," widely discussed by Tarantola (1987) and Stork (1992) \cite{tarantola1987inverse,stork1992reflection}.

The results reinforce the importance of using complementary information—such as well data, velocity inversion, or geological constraints—to reduce interpretive uncertainty in seismic surveys \cite{stork1992reflection,fomel2009absemblance}.

Furthermore, the implementation in Python highlights the potential of open scientific tools in the context of modern computational geophysics \cite{oliphant2007python, lin2012python}. The developed code has a modular structure and can be easily expanded for scenarios with multiple interfaces or lateral velocity heterogeneities, enabling more realistic analyses in future studies.

\bibliographystyle{unsrt}
\bibliography{references}

\end{document}